\documentclass[letterpaper]{article}
\usepackage{isea}
\usepackage[pdftex]{graphicx}
\usepackage{times}
\usepackage{helvet}
\usepackage{courier}
\usepackage[numbers]{natbib}
\usepackage{url}
\usepackage{hyperref}
\usepackage{amsmath} 
\usepackage{graphicx}
\usepackage{float}

\title{Archiving Body Movements: Collective Generation of Chinese Calligraphy}
\author{Aven-Le ZHOU \textsuperscript{1}, Jiayi YE\textsuperscript{2}, Tiancheng LIU\textsuperscript{1}, Kang ZHANG\textsuperscript{1} \\
\textsuperscript{1} The Hong Kong University of Science and Technology (Guangzhou)
\textsuperscript{2} Shanghai Jiao Tong University\\
\texttt{aven.le.zhou@gmail.com}  
}

\begin{document} 

% \makeatletter

% \g@addto@macro\@maketitle{
%     \begin{figure}[H]
%     \includegraphics[width= 15cm]{img/rendering2.jpg}
%     \setlength{\linewidth}{\textwidth}
%     \setlength{\hsize}{\textwidth}
%     \centering
%     \caption{Participant Interacts with the Artwork via Body Movements.\copyright Authors.}
%     \end{figure}
% }
% \makeatother
\maketitle

% \begin{figure}[h]
%     \includegraphics[width= 3.38 in]{img/rendering2.jpg}
%     \caption{Participant Interacts with the Artwork via Body Movements.\copyright Authors.}
%     \end{figure}

\begin{abstract}

As a communication channel, body movements have been widely explored in behavioral studies and Kinesics. Performing and visual arts share the same interests but focus on documenting and representing human body movements, such as dance notation and visual work creation. This paper investigates body movements in oriental calligraphy and how to apply its principles to stimulate and archive body movements. Through an artwork (i.e., Wushu), the authors experiment with an interactive and generative approach to engage the audience’s bodily participation and archive the body movements as a compendium of generated calligraphy. The audience assumes the role of both writers and readers; creating (writing) and appreciating (i.e., reading) the generated calligraphy becomes a cyclical process within this infinite “book,” which aims to motivate further attention and discussion concerning Chinese calligraphy.

\end{abstract}


Embodied Interaction, Body Movement, Bodily Engagement,  Chinese Calligraphy, Interactive Art, AI Art, AI-generated (Painterly) Content.

\section{Introduction}
\label{sec:intro}
Behavioral studies, rooted in the combined fields of sociology and anthropology, have long been interested in human body movement (or motion) \footnote{We can use ``movement'' and ``motion'' interchangeably in the scope of this manuscript, but will continue with ``movement.''} and its meaning. Kinesics, a multidisciplinary nonverbal communication sub-field, deals with interpreting body movements, gestures, and facial expressions \cite{birdwhistell1952introduction}. According to the anthropologist Birdwhistell, who coined the term ``kinesics'' in the 1950s, all body movements have meanings; this nonverbal behavior has a grammar that can be analyzed similarly to spoken language and can vary across different cultures \cite{birdwhistell2010kinesics}.

Creative professionals in performing arts, such as choreographers and dancers, have thoroughly studied how to visualize and document body movements through various explorations since the 18th century \cite{jones2009nama}. Meanwhile, visual artists are more interested in how to interpret and represent the human body movement in various forms of artwork. Similarly, in oriental art and the tradition of calligraphy, body movement has been involved in multiple ways, including its origin, development, and evolution.

Calligraphy creation is a two-part process, ``body movements in the air'' and ``interaction between brush and canvas'' \cite{chiang1973chinese}. The body movements initialize the dynamics and generate force to drive the ink brush and start the creative journey. The contact of brush and canvas, following the driving force and body movement, creates strokes and ink effects. Once the writing begins, these two actions alternate as cause and effect of each other. Calligraphy emphasizes using the body's energy to achieve the artistic creation of ``unity of the body and mind.'' The body movements generate the calligraphy. Meanwhile, the calligraphy and the strokes on the canvas visualize and document the body movement through its principles.

Calligraphy could provide new insights into representing body movements (and archiving them) through its principles; meanwhile, the art could also benefit from these explorations. Our work investigates how to apply the principles of calligraphy to represent and archive body movements. We experiment with an interactive artwork, an infinite and collectively-written ``book'' (i.e., Wushu). It first encourages the audience's bodily engagement through interactive calligraphic visuals (i.e., ink and stroke effects). It then transforms the participants' active participation (i.e., body movements) to Chinese characters using multiple neural networks and then to calligraphic characters, layout, etc. Further, Wushu accumulates participant-generated calligraphy and presents it as a compendium and digital archive, which keeps expanding with the audience's collective participation. 

We highlight our essential contributions in this research investigation as (1) an interactive and generative approach to represent and archive body movements and (2) an artwork and onsite installation to engage the audience's bodily participation and archiving their body movements as a compendium of generated calligraphy.  

\begin{figure*}[t]
    \includegraphics[width=\textwidth]{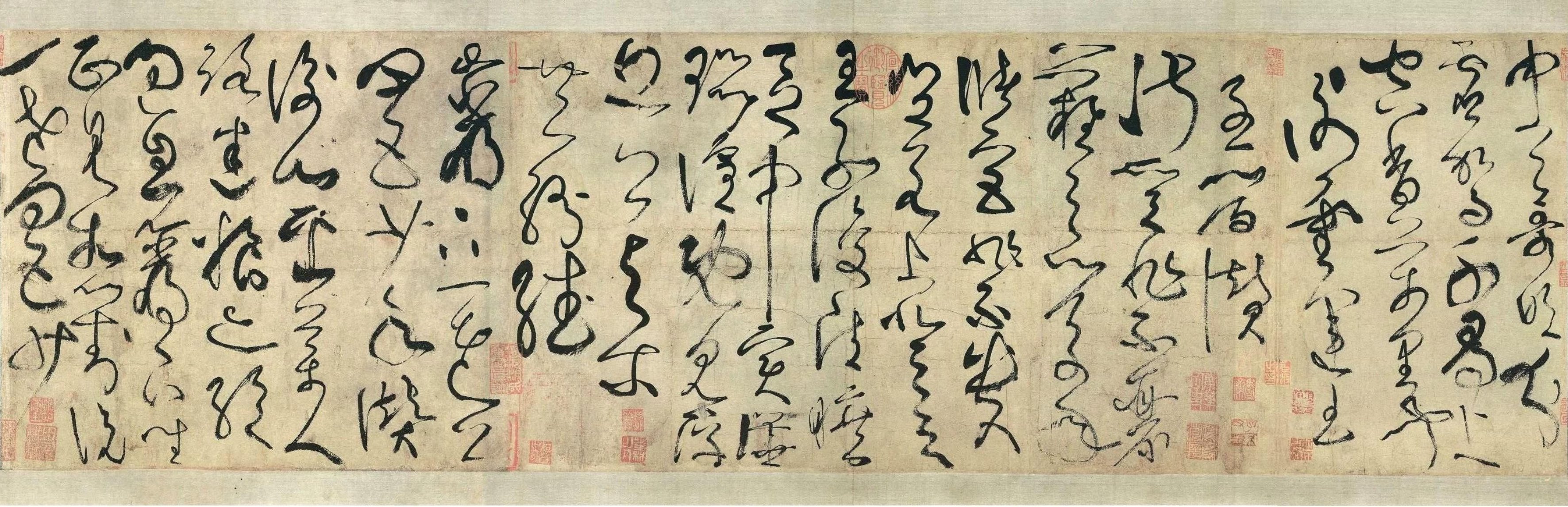}
    \caption{An Example of Zhang Xu's Calligraphy Entitled ``Four Poems.'' \copyright National Palace Museum (CC BY 4.0.)}
    \label{fig:zhangxu}
    \end{figure*}

\section{Contextualization}
\label{sec:Contextualizaiton}

\subsection{Documenting Body Movement in Performing Arts}

Pursuing a tooling function to notate body movements, several systems were invented by performing artists and are widely used for choreography. Labanotation is structured for analyzing and recording body movement with symbols \cite{guest2013labanotation}; similarly, Laban Movement Analysis (LMA) is a method and language for describing, visualizing, interpreting, and documenting movements of the human body \cite{groff1995laban}. Benesh Movement Notation (BMN) records dance movements using symbols representing different body parts \cite{benesh1955benesh}. Eshkol-Wachman Movement Notation (EWMN) treats the body as a stick figure and uses a spherical coordinate system to describe the relationship of different body parts in three-dimensional space \cite{trimingham2017theatre}. 

More scientific and artistic explorations on recording and documenting body movements are not limited to these functional choreographic notations. Stroboscopy, invented by Harold Edgerton, is a technique that uses strobe lights to capture images of moving objects. Edgerton used this technique to record a person jumping rope \cite{belden2011harold}. In his dancing works \textit{Point Point Line} and \textit{Dropping Curves}, William Forsythe explores the relationship between the human body and space by drawing lines and curves through body movement \cite{spier2011william}. Oskar Schlemmer developed the analytic \textit{Stelzenläufer (Slat Dance)}, consisting of dancers wearing stilts and performing movements based on geometric shapes \cite{trimingham2017theatre}. 

\subsection{Representating Body Movement in Visual Arts}

The interest in body movements is not only in performing arts; it is more generally explored as visual art practices beyond merely documentation and archives but as representation. Étienne-Jules Marey coins the term ``Chronophotography,'' and he uses a ``motion capture suit''(i.e., a skeletal marker cloth) and multiple cameras to capture images of moving humans as \textit{Chronophotography of a Walking Man} in the 1890s \cite{braun1992picturing}.

Eadweard Muybridge's photographic study of motion, including the ``Woman Walking Downstairs, Picking up Pitcher, and Turning,'' is a series of instantaneous photographs that document the movement of humans and animals \cite{rossell2013chronophotography}. Frank and Lillian Gilbreth developed a different technique for conducting chronophotographic time-motion studies. They use motion picture cameras to record and analyze workers' movements in repetitive tasks and create a system of 17 behavioral units to describe motion \cite{price2003frank}. 

Like Chronophotography, Marcel Duchamp's \textit{Nu descendant un escalier} (1912) depicts motion by successive superimposed images \cite{duchamp1965marcel}. \textit{Unique Forms of Continuity in Space} (1913) by Umberto Boccioni is a sculpture that explores the relationship between the human body and space \cite{colbert2018umberto}. The sculpture consists of a figure that appears to be moving through space, with its limbs and torso elongated to suggest motion \cite{colbert2018umberto}. In \textit{Forms} (2012), Memo Akten and Davide Quayola explore the relationship between the human body and geometry via algorithm-generated abstract shapes from motion-captured data from athletes \cite{vattano2017dynamic}. 

\begin{figure*}[t]
    \includegraphics[width=\textwidth]{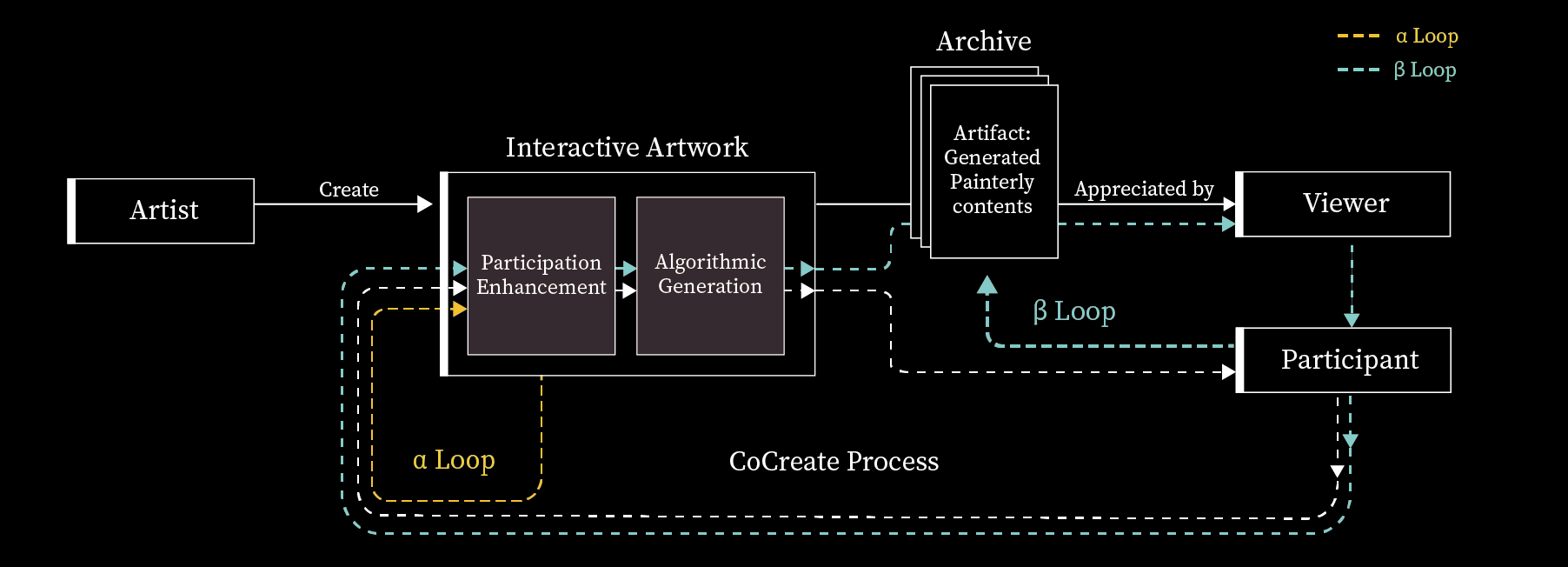}
    \caption{Participant vs. Viewer: the Creative Process. \copyright Authors.}
    \label{fig:creativeprocess}
    \end{figure*}

\subsection{Body Movement in Oriental Calligraphy}

Calligraphy in Eastern Asia refers to writing Chinese characters with ink and brush following different principles, which originated in China and later became popular in both Japan (i.e., shodō) and Korea (i.e., seoye) \cite{da2012chinese}. The Chinese calligraphy chronologically evolves through different levels of body participation that go beyond the common understanding of ``Writing'' and are performative with body involvement \cite{li2010chinese}. The varied bodily engagements with calligraphy result from the changing lifestyle (e.g., furnishing and interior design), various genres of calligraphy, and their corresponding canvas style (i.e., hand scroll, vertical scroll, and multiple-panel). 

Accordingly, Chinese calligraphy starts with writing on a handheld or table scroll, then on the wall or ground (with vertical scrolls) \cite{li2010chinese}. In the late Ming Dynasty, the conventional form of Chinese calligraphy changed from the original horizontal-based hand scrolls to meters-long vertical scrolls. Calligraphers are no longer appropriate to write, either holding the scroll or sitting in front of a table; they begin to stand up and move in front of the wall or on the ground \cite{lin2015printed}, which is naturally accompanied by body involvement. 

Bodily engagement plays a vital role in calligraphy creation and evolution; it further inspires the development of calligraphy. According to a legend story, Zhang Xu created cursive calligraphy (Fig.\ref{fig:zhangxu} shows an example) under the inspiration of Sword Dancing of GongSun. He usually writes when he gets drunk and even uses his hair as a brush with dramatic body movements \cite{barrass2002art}. Huai-su, another master of cursive calligraphy, writes on the wall, where writing becomes dancing with whole-body participation \cite{schlombs1998huai}. In the process of these calligraphy creations, the kinesthetic system is essential; calligraphers coordinate their body movement in writing for artistic expression \cite{chiang1973chinese}. 

Unfortunately, calligraphy is only presented in its static format, and the audience cannot easily imagine the body movements of the calligrapher \cite{marks2011calligraphic}. Through empirical analysis, experienced calligraphers and researchers can identify the ``interaction between brush and canvas'' behind the strokes. They are recognizable since ``the movement is the Calligraphy'' \cite{chiang1973chinese}. However, this requires trained expertise and in-depth knowledge of calligraphy, yet not a complete reflection of body movements. Experts can observe the shape of the strokes, the space between them, and the ink effects to identify the ``interaction between brush and canvas'' but not the ``movements in the air'' that generate them. 

\section{Method}
\subsection{Participant vs. Viewer}

By combining the concepts of interactive art and generative art, we propose a new perspective to examine the typical process of creating and appreciating painterly content (e.g., a painter creates an artwork that is later appreciated by the audience). In interactive art, the audience can be more than a viewer; they can be a participant who engages with the interactive artwork. When the interactive artwork incorporates algorithmic generative capability, especially those algorithms that generate painterly content conditioned on the participant’s real-time inputs, the participant co-creates artifact (i.e., generated painterly content) with the interactive artwork.

Within this paradigm (see Fig.\ref{fig:creativeprocess}), we distinguish the difference between the participant and the viewer (or spectator, bystander, onlooker, etc.) among the audience; the participants gain interactive experiences by engaging with the interactive artwork and become the participant-creator (differs from artist-creator) in the co-creation process, while the viewers do not. During this process, we propose to focus on the mechanisms and design that enhance participant's participation, ideally in a feedback loop of engagement (i.e., the $\alpha$ loop in Fig.\ref{fig:creativeprocess}). Furthermore, we emphasize the distinguished roles of the participant and the viewer, and we should carefully consider them individually; they are interchangeable, and the artifact plays a critical role. The transformation and transition between participant and viewer can be a cyclic process (i.e., the $\beta$ loop in Fig.\ref{fig:creativeprocess}) and improve the audience’s experience with the artwork.

\begin{figure*}[h]
    \includegraphics[width=\textwidth]{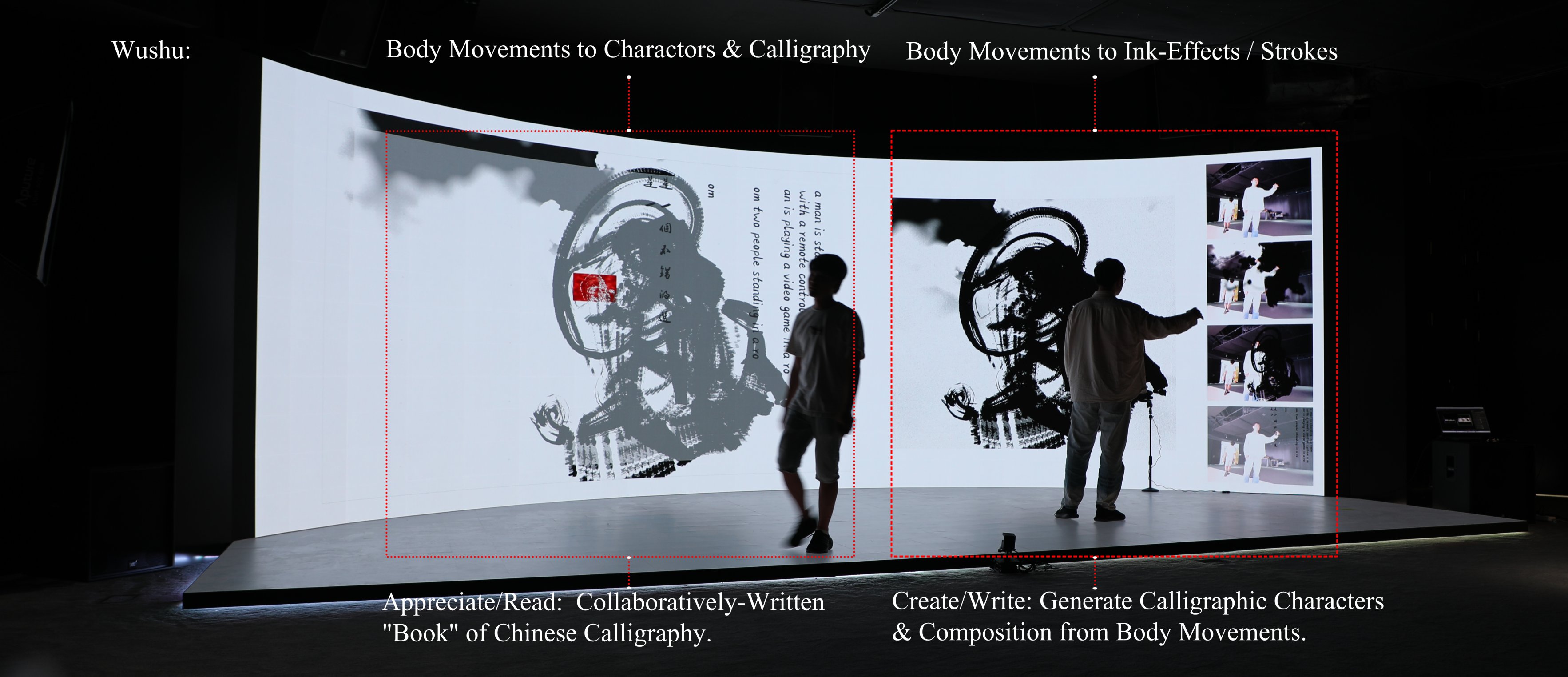}
    \caption{Wushu Installed with a Curved Screen at Computational Media and Arts Lab, HKUST(Guangzhou). \copyright Authors.}
    \label{fig:labscreen}
    \end{figure*}

\begin{figure*}[h]
    \includegraphics[width=\textwidth]{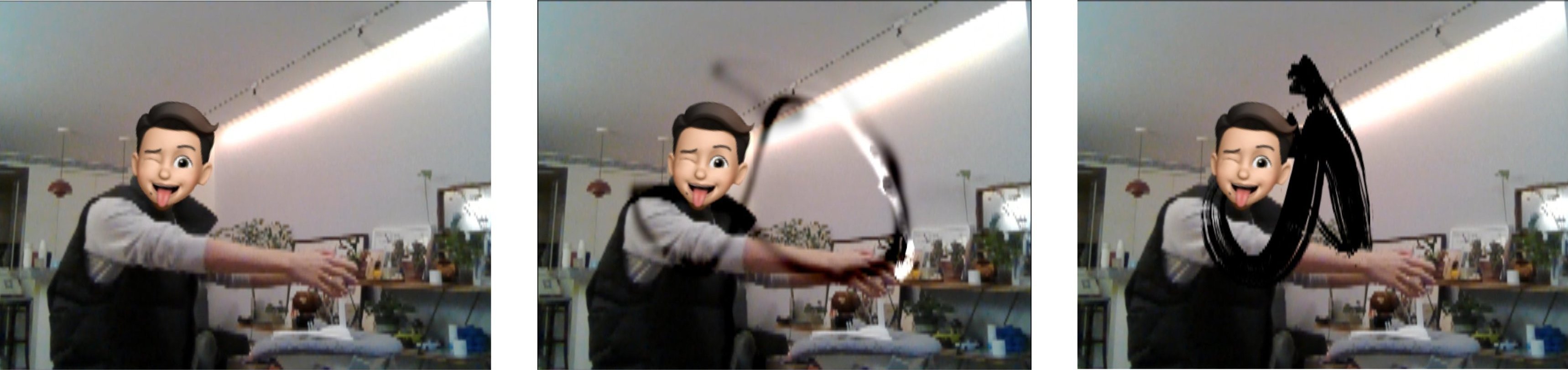}
    \caption{From Body Movements (Left) to Ink Effects (Middle) and Strokes (Right). \copyright Authors.}
    \label{fig:bodymovement}
    \end{figure*}

\subsection{Essential Principles in Calligraphy}
\label{subsec: principles}
From a temporal-spatial perspective, Chinese calligraphy represents sequential strokes: the temporal body movements and the generated moving brushwork and strokes, the spatial structure of the Chinese characters, and their arrangement compositions on the canvas. The brushwork's trajectory is temporal, similar to the musical rhythm and choreography, and spatial, like painting compositions and sculpture structures. The temporal and spatial aspects of calligraphic creation and the references to ``writing'' as calligraphy practice can be summarized as several principles, including ink effect, brush strokes, character composition, and layout composition.

\subsubsection*{Ink Effect and Brush Stroke}

The ink effects of calligraphy depend on the control of the brush and the ink inside; their contact with the canvas with different writing speeds and brush techniques generate various effects. Strokes in calligraphy are about shaping them into diverse expressive forms through brush manipulation. Brush techniques, such as moving forward or backward, fast or slow, lifting or pressing down, turning directions, pausing, etc., produce strokes in different forms that differ in length, roundness, continuity, thickness, and curvature.

\subsubsection*{Character Composition}

The essence of character composition is the manipulation and combination of strokes based on the structure of the original Chinese character. It involves the position and direction of each stroke and the organization of strokes, such as how to cluster them, balance the relationship between them, and control the density of their composition. There are two character composition structures (i.e., Fat vs. Slim) among all calligraphers in different styles, which originate from and refer to the human body description.

\subsubsection*{Layout Composition}

Chinese calligraphy is written from top to bottom and right to left, one column after another. The arrangement at the macro and micro levels reflects the layout composition principle. The primary design concerns are the distance between the characters and the columns, how to cluster them and leave white spaces, and where to place the seals. The characters should not be written to the very edge, and a margin should be left on the four sides. White spaces are expected among columns to allow the energy to flow \cite{chiang1973chinese}.

\section{Wushu: the Infinite Book}
\label{sec:experiment}
We experiment with Wushu to encourage and archive participants' body movements as a collective compendium. The interactive artwork first engages the audience and generates ink effects and strokes from their motion input. Two customized neural networks generate bilingual characters from the audience’s engagements through a ``Motion Caption'' process. Procedural algorithms derived from calligraphy research based on its essential principles manipulate the previously generated characters in custom layout compositions. 

Through the work, the audience participates and automatically “writes” their body movements in the genre of Chinese calligraphy. Each participant does not create independent work but collectively contributes and works on a shared canvas. Only when multiple participant-generated calligraphic characters fully occupy the current canvas will it be automatically archived, and a blank canvas will be initialized. On the other hand, Wushu keeps expanding with the generated calligraphy, which serves as a compendium of the audience's collective creation and a digital archive of their body movements.

\subsection{Audience Engagement and Interaction}

When the audience presents themselves in front of the artwork, Wushu establishes their skeletal body pose and uses the chosen joint (e.g., hand position) and its recorded trajectory to automatically and simultaneously create interactive ink effects and calligraphic strokes. The real-time generated visuals respond to and encourage the participant's bodily engagements and serve as the input for calligraphy generation. 

\subsubsection{Body Pose and Trajectory Estimation}
% \subsubsection{Joint Position Set}
We use Google MediaPipe \cite{lugaresi2019mediapipe} for the human body pose estimation, which provides a set of body joint temporal positions $J_t$ at a fixed period. In equation (1), $i$ denotes the joint index, and $t$ represents the time. 

\begin{equation}
J_t = \{j_t^i\} , 
\end{equation}

Equation (2) shows the joint position (e.g., hand position) $h_t$ can be obtained by selecting the index. We use the current $h_t$ for the real-time control of the ink diffusion effects so that the chosen joint becomes the brush that draws ink effects on the canvas, which responds to the body movements and interaction.

\begin{equation}
h_t = \arg \max_{j_i^t \in \text{hand}} ||j_t^i|| ,
\end{equation}

\begin{equation}
    H_t = \{ h_1, h_2, ... , h_t  \} ,
\end{equation}

The joint trajectory $H_t$ in equation (3) is a sequence that includes all the joint positions from the beginning of the observation. Each $h_i$ in $H_t$ represents the hand position at time $i$ out of $t$. We use $H_t$ for the ink stroke simulation to create the brush strokes that follow the body movement and trajectory. Through the two real-time visual effects (i.e., ink and stroke simulation) feeding back and looping, the artwork attracts the audience to motivate and engage their bodily engagements for further processing. 

\begin{figure}[h]
    \includegraphics[width=3.31in]{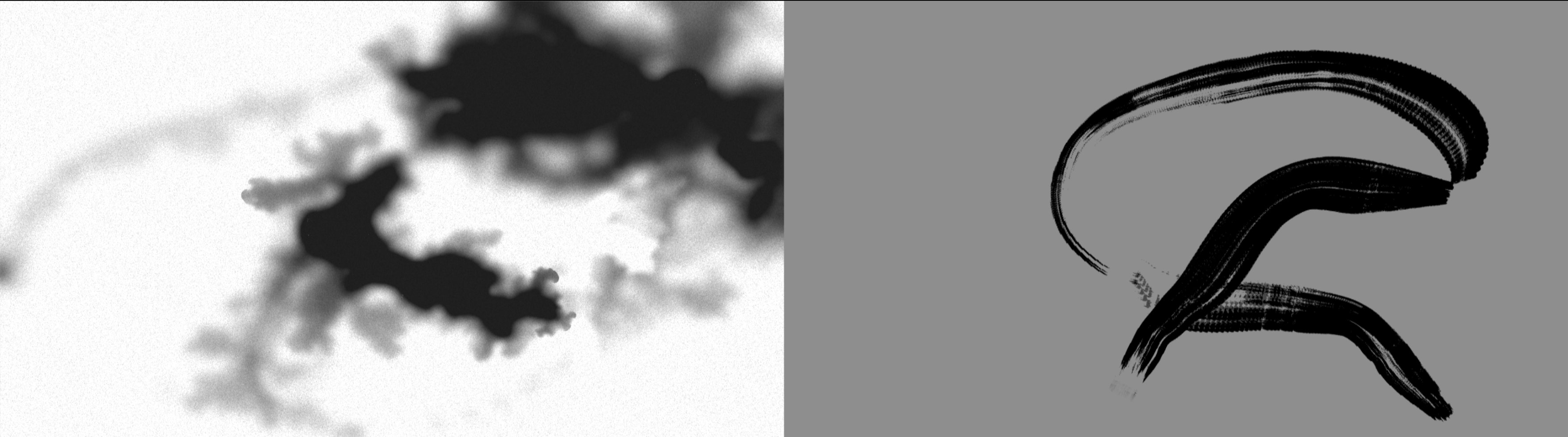}
    \caption{Ink Diffusion Effects (on Left) and Stroke Simulation (on Right) in Wushu. \copyright Authors.}
    \label{fig:effects}
    \end{figure}
    
\subsubsection{Ink Effects and Stroke Simulation}

The temporal evolution of the ink-diffusing texture is simulated based on a standard optical flow algorithm through a partial differential equation (PDE). The optical flow algorithm and partial differential equation (PDE) formula used for the ink diffusion simulation effects:

\begin{equation}
    \frac{\partial T_{t}}{\partial t} = \alpha \nabla^2 T_{t} + \beta \nabla \cdot \left(N(x, y, z, t) + O_t\right) \cdot (1 - T_{t})
\end{equation}

Where $(\frac{\partial T_{t}}{\partial t})$ is the ink texture changing rate; $(\alpha\, \beta)$ are the speed parameters control ink spreading and absorption respectively; $(\nabla^2\, \nabla)$ are the laplacian and gradient operators for diffusion simulation; $(N(x, y, z, t))$ represents Simple3D noise with values in the range [0, 1]; $(O_t)$ is the noise offset at time $(t)$; and $((1 - T_{t}))$ represents the remaining ink portion for simulating absorption.

Based on the input trajectory, we select the joints' convex hull points $ P^a = \{ P_1^a, P_2^a, ... , P_t^a \} = \{ j_{t}^{a} \}$ and $P^b = \{ P_1^b, P_2^b, ... , P_t^b  \} = \{ j_{t}^{b} \}$ to be the two endpoints of lines $ L_t^{ab} $. 

For all the set of points $P^{ab}$, $ \forall P_t^{ab} \in L^{ab} | P_t^{ab} = 0 \lor 255 $ on the line, the color of the individual point in $P^{ab}$ is $Color(p_t)$ denoted as equation (6).  A threshold applies to the color and determines whether it is black or white based on the return value of the noise function. If the noise value is less than or equal to the threshold, the color is black; otherwise, it's white.

\begin{equation}
    L^{ab} = \{ k * P^a + ( 1 - k ) * P^b, 0 <= k <= 1 \} ,
\end{equation}

\begin{equation}
Color(p_t) = 
\begin{cases}
    \text{Black} & \text{if } N(p_x, p_y, p_z) \leq \text{threshold} \\
    \text{White} & \text{if } N(p_x, p_y, p_z) > \text{threshold}
\end{cases}
\end{equation}

In expression of equation (6):
\begin{itemize}
    \item \( Color(p_t) \) represents the color of point \( p \) in the point set \( L^{ab} \) at time \(t \).
\end{itemize}

\begin{itemize}
    \item \( N(p_x, p_y, p_z) \) is the return value from the Simple3D noise function at the coordinates \( (p_x, p_y, p_z) \) of point \( p \).
\end{itemize}

\begin{itemize}
    \item Please note: the noise offset dynamically changes based on signal transformations: \(O_t\).
\end{itemize}

% \begin{itemize}
%     \item The noise offset may dynamically change based on signal transformations: \(O_t\).
% \end{itemize}
% \begin{itemize}
%     \item A threshold value is set to determine whether the color is black or white based on the noise function's output. If the noise value is less than or equal to the threshold, the color is black; otherwise, it's white.
% \end{itemize}

\begin{figure}[h]
    \includegraphics[width=3.38in]{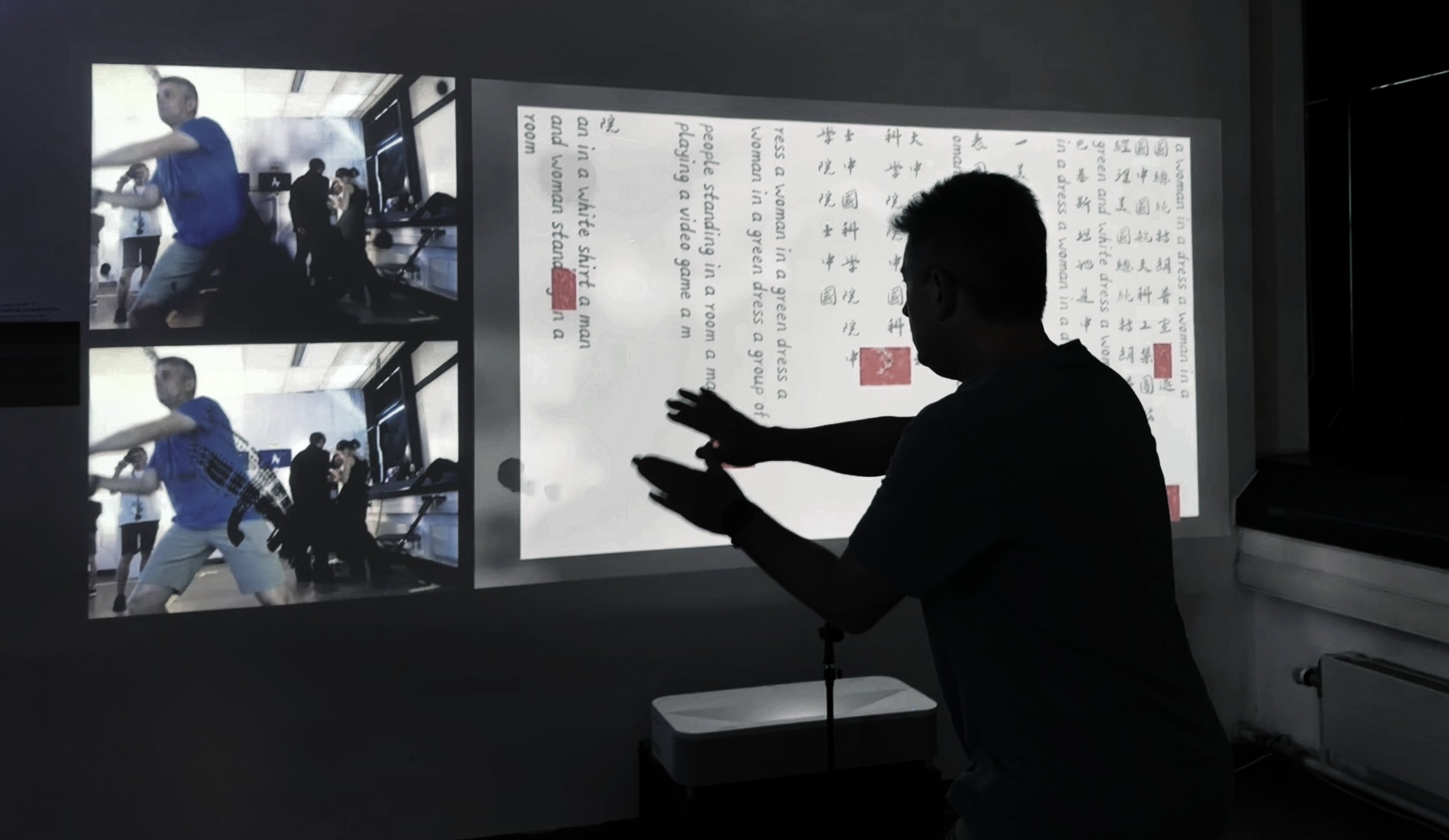}
    \caption{One Participant Interacting with Wushu as if Performing Martial Art at Ars Electronica Festival 2024. \copyright Authors.}
    \label{fig: ars}
    \end{figure}

\subsubsection{Interacion and Input}

The body poses estimation, and the corresponding visual effects enable multiple joints to act as virtual brushes simultaneously. In the iterations of Wushu’s interaction design, we have tried various behaviors as input and control. Using the hand seems most intuitive for most audiences, but it can also be confusing if there is no mechanism to distinguish the start and end of the simulation clearly. Using multiple joints as input simultaneously could work, but only with dancers or other professionals trained with Kinesthetic relevant skills. In the open-to-public exhibition, Wushu asks the audience to close their two wrists to start the simulation (and then open their wrists to stop the simulation), which makes the participants look like they are performing martial arts when interacting with the work (see Fig.\ref{fig: ars}).

\begin{figure*}[t]
    \includegraphics[width=\textwidth]{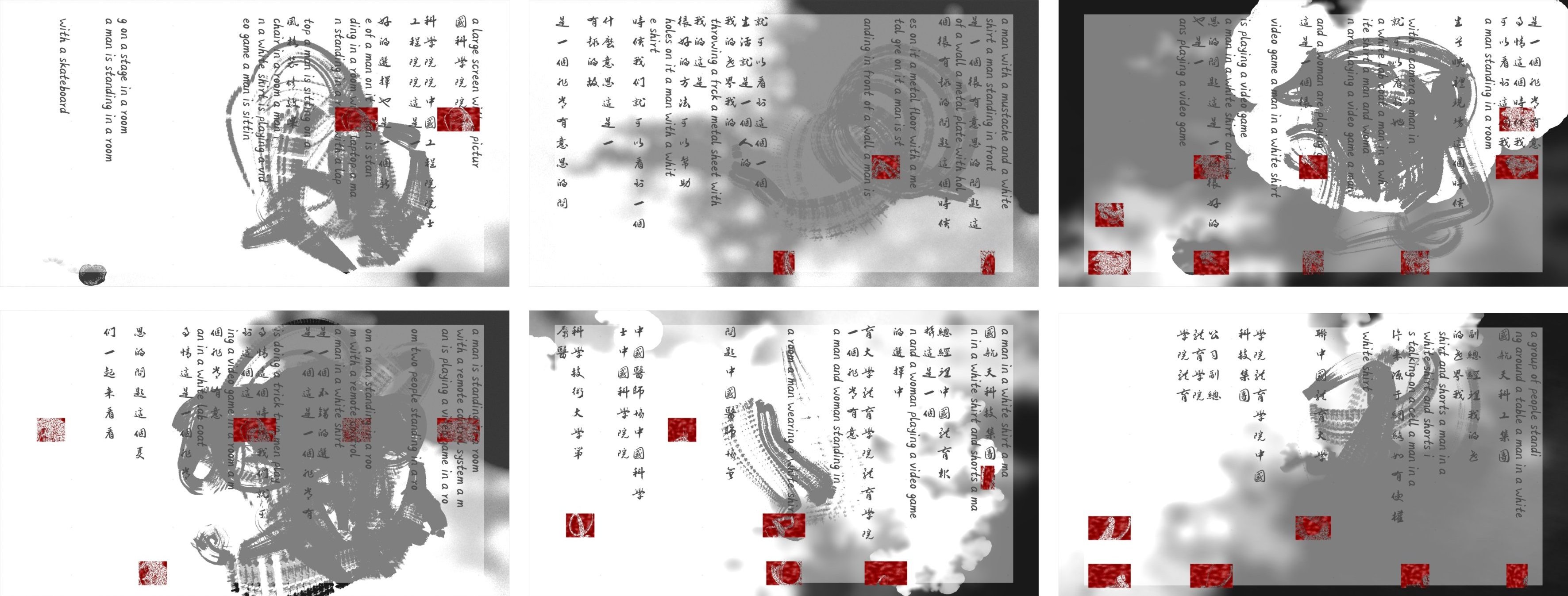}
    \caption{Sample Pages of the Compendium. 
    \label{fig:compendium}
    \copyright Authors.}
    \end{figure*}

\subsection{Participatory Generation of Calligraphy}

Through real-time generated visual effects, Wushu can stimulate participants to engage with more dynamic body movements when interacting with the work. Wushu subsequently and automatically “writes” the participant's body movements in the Chinese calligraphy genre by generating characters from the live camera images, manipulating and arranging them as calligraphic characters and layouts. 

\subsubsection{Motion Caption to Generate Characters} 

We integrate two image caption (image-to-text) algorithms, BLIP and vit-gpt2, for English and Chinese captions to generate characters (i.e., textual descriptions) from the body movements. These two models are based on visual language pre-training (VLP) and translate captured body movement images into textual descriptions. The BLIP model specializes in visual language understanding and generating tasks and can process visual data to produce descriptive notes in English. The ViT-GPT2 model, for Chinese language processing, combines image encoding and GPT-2 decoding using Google's Visual Transformer (ViT) to generate the corresponding Chinese captions.

The two image-to-text models provide semantic descriptions of the body movement images in English and Chinese. We especially incorporate bilingual captions to broaden the context and make the work accessible to audiences who do not understand Chinese. With the real-time captured images as input, the process can be denoted as: \[ T = f(I) \]
Where \( T \) is the set of textual descriptions corresponding to the images in set \( I \), processed by the function \( f \), which encompasses the image-to-text algorithms.

\subsubsection{Characters Composition}

We work with a professional calligrapher to conduct an empirical analysis to develop a custom algorithm to manipulate and simulate character composition and layout (see the results in Fig.\Ref{fig:compendium}). The custom algorithm controls the dimensions of the characters in terms of height and width, which alters the aspect ratio of the character and results in various appearances, e.g., tall or short, fat or slim. 

The character and layout composition simulation starts with applying a calligraphic font to the image caption texts and obtaining the calligraphic characters, denoted as \( C = \{C_1, C_2, ..., C_n\} \). The algorithm alters each character's aspect ratio by function \( f \), which reflects on the nuances of the captured body movements. 

It takes the recorded trajectory of the body joint as input and calculates the bounding box to determine the ratio of the character(s). The calligraphic presentation of character composition based on the body movements and its textual caption can be summarized as \[ M(C) = \{ f(C_1), f(C_2), ..., f(C_n) \} \]

\subsubsection{Layout Composition}

We use randomized grouping to generate white space for layout composition. When initializing the canvas, the algorithm generates a new layout by grouping the columns with calligraphic characters and inserting the white space (i.e., empty columns) between the groups. The pre-determined range of the grouping and the spacing between groups ensure the calligraphic aesthetic and keep consistency when introducing randomness (see Fig.\ref{fig:compendium}).

Both the grouping and white spaces are variational within a dynamic range to increase the visual rhythm of the composition and let the "energy" flow. Consider a set of columns \( L = \{L_1, L_2, ..., L_n\} \). These columns are grouped into \( G = \{G_1, G_2, ..., G_m\} \). Intra-group spacing is fixed at \( s_{intra} \), while inter-group spacing varies within \( s_{inter} \in [S_{min}, S_{max}]\). In our work, the layout composition based on the calligraphic rules can be shown as follows:
\[ L(G, s_{intra}, s_{inter}) = \bigcup_{i=1}^{m} \left( G_i, s_{intra}, s_{inter} \right) \]

\subsection{The Compendium, Archive, and Installation}

When interacting with Wushu, the participant generates calligraphy and contributes to the collective efforts by working on a shared canvas with other participants asynchronously. In this way, the current participant collaborates with previous and future ones to complete the calligraphy piece. When the participatory-generated calligraphic contents fill the canvas, Wushu automatically archives them (i.e., saves them to the compendium). Wushu keeps expanding with the generated calligraphy and becomes a compendium of the audience’s collective creation, which is also the archive of all their body movements (see examples in Fig.\Ref{fig:compendium}).

Wushu's installation, dimensions, and interface design are flexible. Fig.\ref{fig:bodymovement} shows our first test installation with a big curved screen open to the internal audience for participation. Our simplified installation setup at Ars Electronica Campus Exhibition used a portable 4K projector as the interactive screen (in Fig.\Ref{fig: ars}). Both worked smoothly; we conclude the essence of the installation is, hardware-wise, to keep the screen big enough and ideally identical to the human scale so that a light touch of immersion could enhance the interactive experience. 

\section{Related Work}

In the \nameref{sec:Contextualizaiton} Section, we have reviewed and discussed the interests of documenting and representing body movements in performance and visual art. Many researchers and artists investigate the imaginative and unseen relationships between body movements and visual forms through various analytic techniques and creative practices. We discover that pioneer calligraphers have focused on better transforming the body's energy and movements to calligraphy practice and pushing the boundary of the art tradition. However, there is a lack of work on linking body movements with calligraphy characters and archiving them.

\subsection{From Body Movements to Forms} 

Several visual artworks explore the relationship between body movements and visual forms. For example, Tobias Gremmler, in his ``Kung Fu Motion Visualization Series (2016),'' explores how ``velocity transforms into matter.'' He displays particles whose opacity is proportional to the speed of the performer's body movements, while stationary body parts remain invisible. In his series of daily computational sketches, Zachary Lieberman interprets human body movements with simple yet elegant visual forms. However, these works focus only on the relevance of body movements to abstract visuals and do not consider the connection between body movements and calligraphy characters, which is the main topic of our research.

\subsection{Modern Calligraphy Practices}

Modern pioneer calligraphers in Eastern Asia have recently focused on large-scale works with a single or a few characters on the canvas. Calligraphers need to work through full-body engagements to write big-size characters. Japanese artist Yuichi Inoue, known as ``the ghost of calligraphy,'' often stands on the canvas barefoot and writes to transform the body movements and energy to the work. 

Similarly, Chinese contemporary calligrapher Dongling Wang features huge-size characters and wild cursive calligraphy. He summarizes that the whole-body writing practice relies on ``the strength of waist, endurance, agile body, and its movements, as well as their overall balance.'' In the advancement of calligraphy art practice, exploring body movements is still avant-garde.

\subsection{Interactions with (Calligraphic) Characters} 

Other interactive artworks investigate the relationship between body movement and characters and calligraphy. For example, Google Creative Lab collaborates with Bill Jones to design interactions based on body movements. The experiment invites users to explore their body movements and connect with Bill Jones' iconic dance, using the PoseNet machine learning model to interact with the speech-to-text characters. 

In his ``Movement in Time, Part 2 (2016),'' Bryan Chung applies computer vision techniques to speculate the transformation of body movements in martial arts to cursive calligraphy. The body movements are mapped to a given thousand characters through machine-learning clustering algorithms. These works create real-time experiences and connect body movements to calligraphic characters but do not present them as calligraphy or archive them, which identifies the research gap for our work.

\section{Conclusion and Discussion}

On archiving body movements, we experiment with Wushu -- an interactive artwork that generates calligraphy from body movements. We invite the audience to participate as the co-creator of the ``book,'' a digital compendium of the collective calligraphic works and an archive of the body movements. Our approach establishes a new calligraphy creation process accessible to individuals regardless of their prior knowledge of the traditional art form. 

Wushu also serves as a virtual platform to review and appreciate the audience-generated works; the audience assumes the role of both writers and readers in the context of the ``book.'' Creating (writing) and appreciating (reading) the generated calligraphy becomes a cyclical process within this infinite ``book.'' Wushu can stimulate further attention and discussions concerning Chinese characters and calligraphy, such as their visual forms, linguistic meanings, and cultural essence. 

\section{Limitations and Future Work}

However, Wushu has its limitations, challenges, and opportunities that we would like to address in the future. The compendium is an essential component of Wushu's concept and the archive. The audience finds it attractive that their collective efforts will form an actual ``book.'' However, we have not yet figured out the best format to present the archive, except for the real-time showcasing as part of Wushu's interactive experiences. We currently save the digital pages as image files. Still, we are considering an online gallery that updates in real-time and publishes the archive publicly online or a physical book in print that mimics the reading experience. We will further explore and investigate how to present the archive better in the future.

During the exhibition of Wushu, we notice a significant difference in the performance of the English and Chinese caption models. The English caption model we acquired generates more accurate and fluent captions than the Chinese one. We attributed this difference not to the lack of AI research on Chinese captioning but to the scarcity and quality of the data sources. The large models that we used for captioning were trained with semantic information from various online data, while the predominant source of Chinese texts is of lower quality and diversity than the English data. Although this manuscript and research do not address the data and AI bias issues, we intend to explore them further.

\section{Acknowledgments}
Aven Le Zhou created the artwork discussed in this manuscript, encompassing its conceptual framework and technical development. He authored the major part of the manuscript unless specifically attributed otherwise. The artwork was shown and documented in various venues, such as the Ars Electronica Festival 2024, CVPR 2024 AI Art track and Digital Art China 2024. Jiayi Ye contributed insights into broader calligraphy practice and authored the calligraphy-related sections, including the \nameref{subsec: principles} and parts of the \nameref{sec:intro}. Tianchen Liu provided expertise in technical expressions for the \nameref{sec:experiment}. Kang Zhang guided the overall project and manuscript development as the PhD supervisor.

% \section{References}
% The title “References” should be 12 point, bold style, centered. Editorial standards adhere to the guidelines set by the Chicago Manual of Style, 17th ed. References should be 9 point, regular type. List them in numerical order immediately after your essay. The numbers should be cross-referenced within the essay, with numbers placed at the end of the sentence in square brackets, with a space after the full stop, as shown at the end of this sentence \cite{boden92}.
% Can we add another new reference? \cite{ritchie07}

\bibliographystyle{isea}
\bibliography{isea}
\section{Authors Biographies}
\subsubsection{Aven Le ZHOU} (b. 1991, Wuhan, China) is an artist scholar and Ph.D. researcher in Computational Media and Arts at The Hong Kong University of Science and Technology (Guangzhou). He is the chair of Leonardo LASER talk (Guangzhou), an elected Affiliated Fellow of the Chartered Society of Designers, and a Fellow of the Royal Society of Arts since 2022. He has held an assistant professorship at Xi'an Jiaotong-Liverpool University since 2020. Prior to that, he was a Clinical Instructor (2019 - 2020) and Research Fellow (2014 - 2019) at Interactive Media Arts at New York University Shanghai.

\subsubsection{Jiayi YE} (b. 1989 in Shanghai, China) is a calligrapher and researcher. She is primarily focused on digital calligraphy art, cross-cultural communication related to Chinese calligraphy, and the aesthetics of this ancient art form in the modern context.

\subsubsection{Tiancheng LIU} (b. 1996, China) is a computer science researcher currently focusing on computing aesthetics and Chinese calligraphy. He was a researcher and engineer in cloud computing and networking, and intelligent transportation IoT.

\subsubsection{Kang ZHANG} is Acting Head and Professor of Computational Media and Arts, Hong Kong University of Science and Technology (Guangzhou), Professor of Division of Emerging Interdisciplinary Areas, HKUST, Professor Emeritus of Computer Science, The University of Texas at Dallas, and Guest Professor of China Academy of Arts. He was a Fulbright Distinguished Chair and among 2023 Stanford's List of Top 2\% Scientists (in Career).

\end{document}